\documentclass{iopart}
\usepackage{iopams}
\expandafter\let\csname equation*\endcsname=\relax
\expandafter\let\csname endequation*\endcsname=\relax
\usepackage{rotating}
\usepackage{amsfonts}
\usepackage{amssymb}
\usepackage{amsmath}
\usepackage{latexsym}
\usepackage{graphicx}
\usepackage{cite}
\usepackage{multirow}
\usepackage[english]{babel}
\usepackage{xspace}

\newcommand{\Htp}{\ensuremath{\rm H_2^+}\xspace}
\newcommand{\Dtp}{\ensuremath{\rm D_2^+}\xspace}

\begin{document}

\title{\Htp in a weak magnetic field}

\author{H\'ector Medel Cobaxin}
\address{Universit\'e de Reims Champagne-Ardenne,
 Groupe de Spectrom\'etrie Mol\'eculaire
 et Atmosph\'erique (UMR CNRS 7331),
 U.F.R. Sciences Exactes et Naturelles,
 Moulin de la Housse B.P. 1039,F-51687 Reims Cedex 2, France}
\ead{hector.medel-cobaxin@univ-reims.fr}
\author{Alexander Alijah}
 \address{Universit\'e de Reims Champagne-Ardenne,
 Groupe de Spectrom\'etrie Mol\'eculaire
 et Atmosph\'erique (UMR CNRS 7331),
 U.F.R. Sciences Exactes et Naturelles,
 Moulin de la Housse B.P. 1039,F-51687 Reims Cedex 2, France}
\ead{alexander.alijah@univ-reims.fr}

\author{Juan Carlos L\'opez Vieyra} 
\address{Instituto de Ciencias Nucleares, Universidad Nacional
  Aut\'onoma de M\'exico, Apartado Postal 70-543, 04510 M\'exico,
  D.F., Mexico}
\ead{vieyra@nucleares.unam.mx}

\author{Alexander V. Turbiner} 
\address{Instituto de Ciencias Nucleares, Universidad Nacional
  Aut\'onoma de M\'exico, Apartado Postal 70-543, 04510 M\'exico,
  D.F., Mexico}
\ead{turbiner@nucleares.unam.mx}

\begin{abstract}
  The electronic energy of \Htp in magnetic fields of up to $B=0.2
  B_0$ (or 4.7 $\times 10^4$ Tesla) is investigated.  Numerical values
  of the magnetic susceptibility for both the diamagnetic and
  paramagnetic contributions are reported for arbitrary orientations
  of the molecule in the magnetic field.  It is shown that both
  diamagnetic and paramagnetic susceptibilities grow with inclination,
  while paramagnetic susceptibility is systematically much smaller
  than the diamagnetic one.  Accurate two-dimensional Born-Oppenheimer
  surfaces are obtained with special trial functions. Using these
  surfaces, vibrational and rotational states are computed and
  analysed for the isotopologues \Htp and \Dtp.
\end{abstract}

\pacs{31.15.Pf,33.15.Kr,72.20.-g,33.15.Hp,33.20.Vq}
\vspace{2pc}

\noindent{\it Keywords\/}: Variational method, weak magnetic fields,
magnetic susceptibility, rovibrational states

\submitto{\jpb}
\maketitle

\section{Introduction}

Since the pioneering work of de Melo {\em et al.}~\cite{Melo:76676} on
the molecular ion \Htp in strong magnetic fields, $B\geq B_0$
($B_0=2.35\times10^9$ Gauss $=2.35\times10^5$~T), many studies have
been conducted for this system under such conditions (see for
example~\cite{Kappes:954542}-\cite{Song:13064305} and references
therein) where the electronic energy of the ground and first excited
states as well as some rotation-vibrational states have been studied.
\Htp has been used as a test system for the investigation of the
validity of approximations commonly made in field-free molecular
physics, such as the Born-Oppenheimer
approximation~\cite{Schmelcher:88672,Schmelcher:886066}.  Though \Htp
can be considered a benchmark molecule for the development of
appropriate theoretical methods for the accurate computation of
molecular structure and properties in magnetic fields that may be
extended to more complex systems~\cite{Turbiner:06309,Turbiner:07267,
  Turbiner:07053408, Turbiner:10042503}, only few studies have been
reported in the range of {\em small} fields ($0\leq B \leq
B_0$)~\cite{Wille:87L417,Larsen:821295}, where electronic energies of
the ground state and physical features such as a qualitative evolution
of the rotational levels as function of the field are presented.

The goal of the present study is to investigate the electronic ground
state of the molecular system \Htp placed in a weak magnetic field,
$B\leq 0.2 B_0$. Overall in this domain of field strength, the effects
of the magnetic field cannot be treated accurately via perturbation
theory. In the first part of the present work we use physically
motivated, specially tailored trial
functions~\cite{Turbiner:03012504,Turbiner:04053413,Turbiner:06309} to
obtain sufficiently accurate estimates of the electronic energy over a
range of field strengths up to $B=B_0$ and different inclinations of
the molecular axis with respect to the field direction. In the second
part we investigate the vibrational and rotational structure of \Htp
and \Dtp in the external magnetic field.

\section{Hamiltonian}
\begin{figure}[htb]
\centering
\includegraphics[angle=0, width=0.40\textwidth]{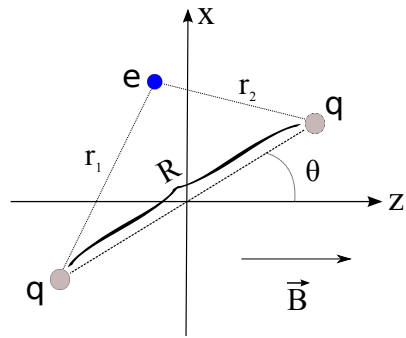}
\caption{The homonuclear molecular ion consisting of two centers of
  charge $q$ separated by a relative distance $R$, and one electron
  $e$. The system is placed in a uniform magnetic field $\vec{B}$
  along $z$-axis.}
\label{htpfig}
\end{figure}

We consider a homonuclear molecular ion formed by two nuclei of charge
$q$ separated by a distance $R$, and one electron $e$ placed in a
uniform magnetic field $\vec{B}$ oriented along the $z$-axis.  The
reference point for coordinates is chosen to be at the midpoint of the
line connecting the nuclei which in turn forms an angle $\theta$ with
respect to the magnetic field direction (see Figure \ref{htpfig}).
In the framework of non-relativistic quantum mechanics, {\em i.e.}
neglecting spin interactions, following pseudoseparation of the center
of mass motion~\footnote{For further information see
  \cite{Avron:78431,Kappes:954542,Schmelcher:88672,Schmelcher:886066}}
and resorting to the Born-Oppenheimer approximation of order zero,
{\em i.e.} neglecting terms of order smaller than $1/\cal{M}$
($\cal{M}$ is the total mass of the system), the Hamiltonian that
describes the system is given by
\begin{equation}
\label{AHamil}
\hat{\cal H}= \frac{2}{M_s}\left[\hat{P}_R-\frac{q}{2}\vec{\cal A}_R\right]^2 + \frac{1}{2m_e}\left[\hat{p}-e\vec{A}_e\right]^2+\frac{q^2}{R}+q e\left(\frac{1}{r_1}+\frac{1}{r_2}\right),
\end{equation}
where $M_s$ is the total mass of the nuclei, $q$ is the nuclear
charge, $\hat{P}_R = -i \hbar \nabla_{R}$ is the momentum operator and
$\vec{A}_R$ ($\vec{A}_R=\frac{1}{2}\,\vec{B}\times \vec{R}$) is the
vector potential for the relative motion ${\vec R}$ of the nuclei; $e$
and $m_e$ are the electronic charge and mass, respectively; $\hat{p}=
-i \hbar \nabla_{r}$ and $\vec{A}_e$ are the momentum operator and
vector potential for the electron which is at the position $\vec{r} =(x,y,z)$; $r_1$ and
$r_2$ are the distances between the electron and each of the
nuclei. In the Hamiltonian~\eqref{AHamil} the first term is the
kinetic energy of the nuclear relative motion in a magnetic field; the
remaining terms correspond to the electronic Hamiltonian, written in
the Born-Oppenheimer approximation of zero order.  In the remainder of
this article, atomic units shall be used, {\em i.e.} distances are
measured in Bohr, $a_0=1\,$ a.u., energies in Hartrees, $E_h=1\,$
a.u. and $\hbar=|e|=m_e=1$.

\section{Solving the electronic Schr\"odinger equation}

To solve the electronic Schr\"odinger equation an appropriate gauge
for $\vec{A}_e$ must be chosen. Though the problem is in principle
gauge invariant this is not the case if the equation is solved
approximately~\cite{Larsen:821295,Kobe:833710,Larsen:07042502}.  We
have therefore adopted the strategy of introducing a variational
parameter, $\xi$, in the definition of the gauge which is then varied
together with the variational parameters of the wave function. For a
magnetic field directed along the $z$-axis $\vec{B}=B\hat{z}$, a
suitable vector potential is
\begin{equation}
\label{gengauge}
\vec{A}_e=B\left[(\xi-1)y,\xi x,0\right],
\end{equation}
%
where $\xi$ is the parameter of the family of Coulomb gauges. With
$\xi=0$ the Landau gauge is obtained, while $\xi=1/2$ corresponds to
the symmetric gauge.  Substituting \eqref{gengauge} into
\eqref{AHamil} we obtain the electronic Hamiltonian (the last three
terms in \eqref{AHamil}) in the form
\begin{eqnarray}
\label{ElectHamil}
\hat{\cal H}_{elec}=&-&\frac{1}{2}\nabla^2-iB\left[ (\xi-1)y \partial_x+ \xi x \partial_y\right]  \\\nonumber
 &+&\frac{1}{2}B^2\left[\xi^2x^2+(1-\xi)^2y^2 \right]+\frac{q^2}{R}-\left(\frac{1}{r_1}+\frac{1}{r_2}\right).
\end{eqnarray}
As usual, in (\ref{ElectHamil}) the contribution to the energy due to
the Coulomb interaction between the nuclei, {\em i.e.} $q^2/R$, is
treated classically.  Hence, $R$ is considered an external parameter.
\subsection{Trial functions}\label{trialf}
A set of physically adequate real trial functions introduced in
\cite{Turbiner:03012504,Turbiner:04053413,Turbiner:06309} are used to
calculate the total energy of the electronic Hamiltonian
(\ref{ElectHamil}). Thus, the trial function employed in the present
study is a linear superposition of three particular functions,
\begin{equation}\label{Trialf}
\Psi=A_1\Psi_1+A_2\Psi_2+A_3\Psi_3\,,
\end{equation}
where
\begin{equation}
\label{psitotal}
\Psi_1= \ e^{ - \alpha_1 (r_1+r_2)}e^{-B[\beta_{1x}\xi x^2 +\beta_{1y}(1-\xi)y^2]}\,,
\end{equation}
is a Heitler-London type function,
\begin{equation}
\Psi_2= \left(e^{ - \alpha_2 r_1}+e^{ - \alpha_2 r_2}\right)e^{-B[\beta_{2x}\xi x^2 +\beta_{2y}(1-\xi)y^2]}\,,
\end{equation}
is a Hund-Mulliken type function, and
\begin{equation}
\Psi_3= \left(e^{ - \alpha_3 r_1-\alpha_4 r_2}+e^{ - \alpha_3 r_2-\alpha_4 r_1}\right)e^{-B[\beta_{3x}\xi x^2 +\beta_{3y}(1-\xi)y^2]}\,,
\end{equation}
is a Guillemin-Zener type function, all multiplied with exponential
terms that correspond to the lowest Landau orbital.

Without loss of generality one of the linear parameters $A_{1,2,3}$
may be set equal to one, hence the trial function consists of 13
variational parameters.  For the parallel configuration
the parameters are not independent and must obey the symmetry
relations $\beta_{1x}=\beta_{1y}$, $\beta_{2x}=\beta_{2y}$ and
$\beta_{3x}=\beta_{3y}$, reducing the number of variational parameters
to ten. The trial function~\eqref{Trialf} defined in this way is
expected to provide an accurate approximation to the exact electronic
wave function of the ground state of molecular ion \Htp for a large
variety of strengths and inclinations of the magnetic field.

Calculations are performed using the minimization package MINUIT from
CERN-LIB. Numerical integrations were done with a relative accuracy of
$\sim 10^{-6}$ using the adaptive NAG-LIB (D01FCF) routine.

\subsection{Results}
Using the trial function (\ref{psitotal}) presented in Section
\ref{trialf}, two-dimensional potential energy surfaces of the
electronic energy have been obtained variationally as function of the
internuclear distance, $R$, and the inclination $\theta$ (see Figure
\ref{htpfig}).

\begin{figure}[htp!]
\centering
\includegraphics*[angle=-90, width=0.85\textwidth]{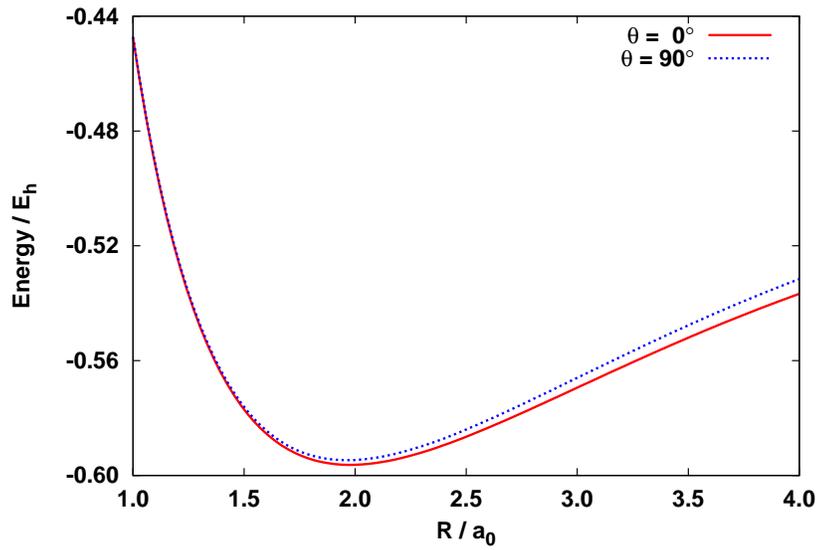}
\caption{Potential energy curves for \Htp in an external field of $B=0.2 B_0$ at
parallel ($\theta=0^{\circ}$) and perpendicular ($\theta=90^{\circ}$) 
orientations.}
\label{B02Plots}
\end{figure}

\begin{figure}[htp!]
\centering
\includegraphics*[angle=-90, width=0.85\textwidth]{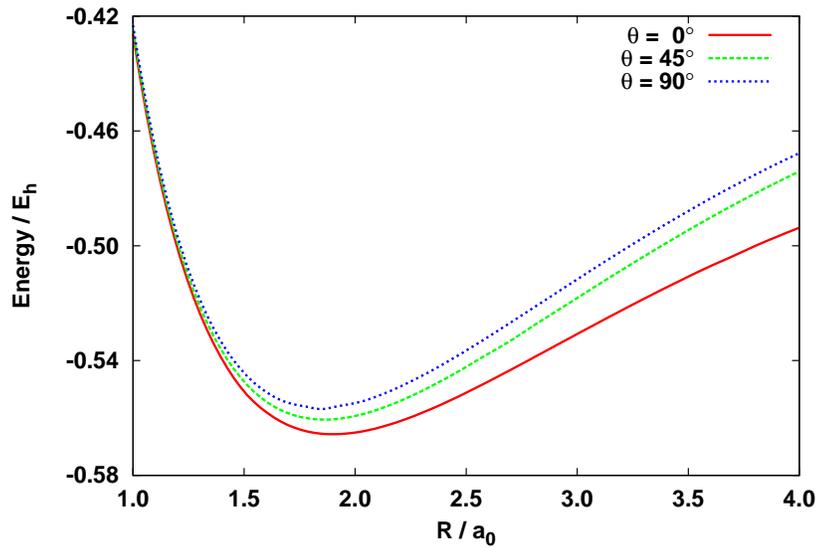}
\caption{Potential energy curves for \Htp in an external field of $B=0.5 B_0$ 
at orientation angles $\theta=0^{\circ}$, $\theta=45^{\circ}$
and $\theta=90^{\circ}$.}
\label{B05Plots}
\end{figure}

\begin{figure}[htp!]
\centering
\includegraphics*[angle=-90, width=0.85\textwidth]{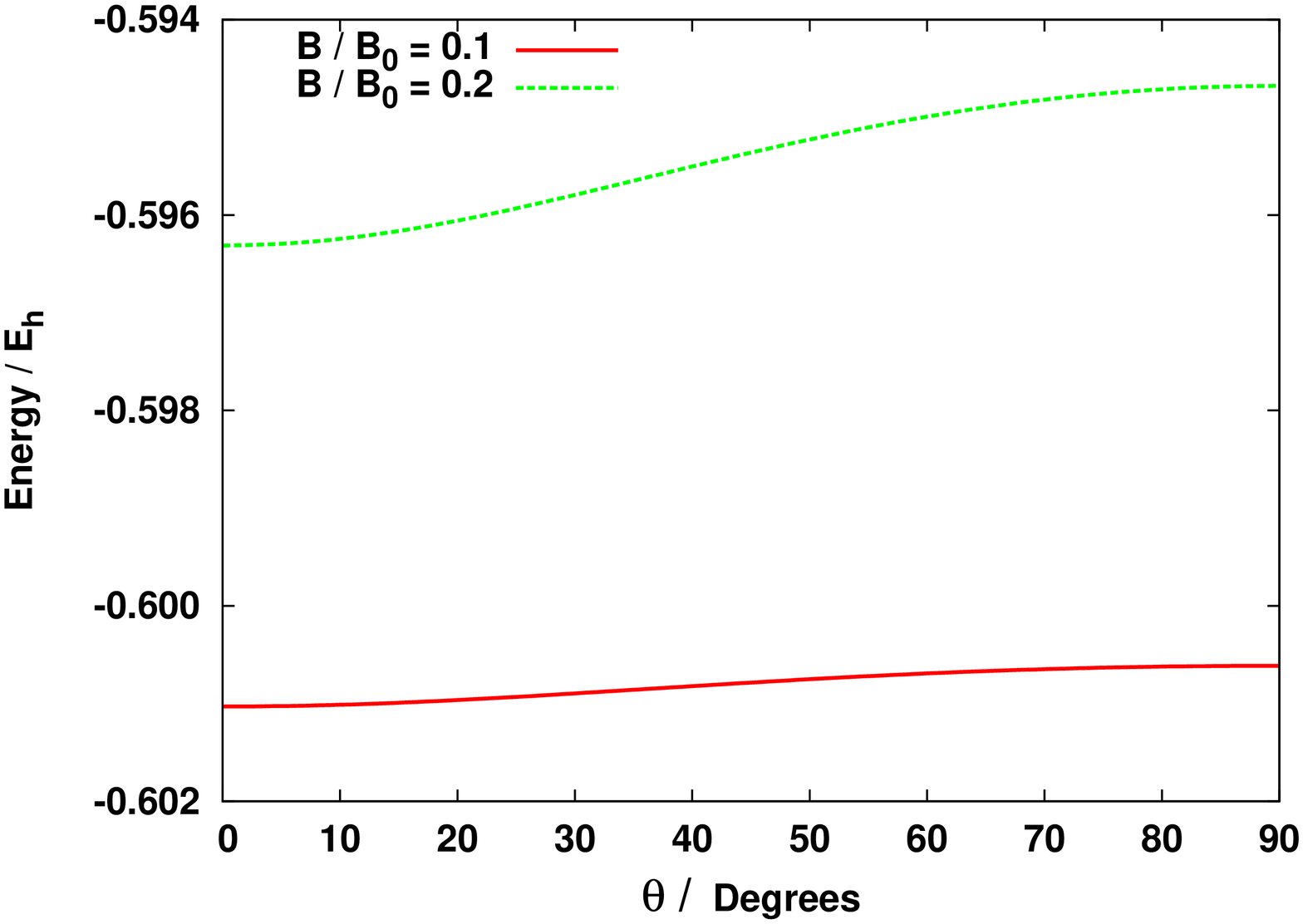}
\caption{Electronic energy at equilibrium distance as function of
  inclination for \Htp in an external field of $B=0.1 \, B_0$ and $0.2 \, B_0$.}
\label{h2pET}
\end{figure}

\begin{figure}[htp!]
\centering
\includegraphics*[angle=-90, width=0.85\textwidth]{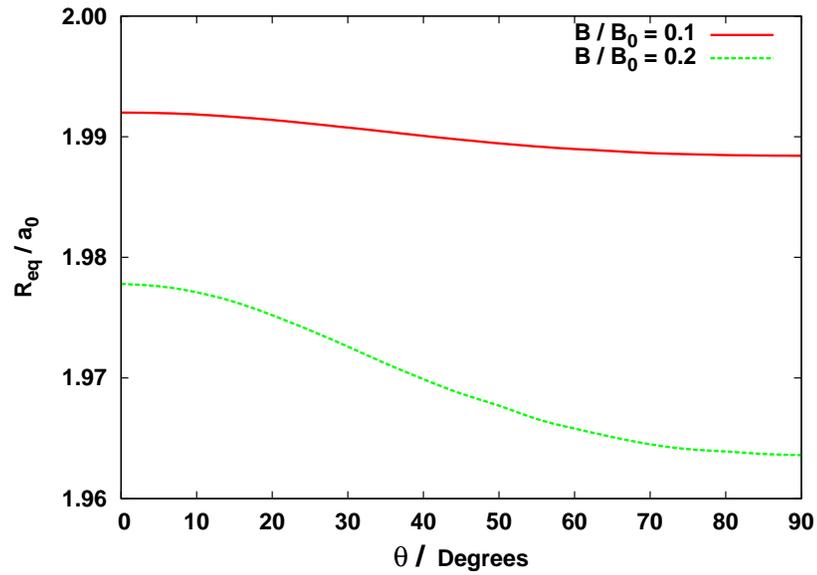}
\caption{Equilibrium distance in a.u. of \Htp in an external field as function of
  the inclination angle and field strengths of $B=0.1 \, B_0$ and $0.2 \, B_0$.}
\label{h2pRT}
\end{figure}

As examples we show in Figures~\ref{B02Plots} and \ref{B05Plots}
sections of the potential surface at different inclinations,
$\theta=0^{\circ}, 45^{\circ}$ and $90^{\circ}$, for $B=0.2 B_0$ and
$B=0.5 B_0$.
The most stable configuration is achieved for parallel orientation of
the molecule, which is a well known result for $B\geq
10^5\,$T~\cite{Turbiner:06309}. At perpendicular orientation an energy
ridge shows up which can be interpreted as barrier of a hindered
rotation. This fact is shown in more detail in Figure~\ref{h2pET}
where the electronic energy is plotted as function of the inclination
for the fields $B =0.1\, B_0$ and $0.2 \, B_0$. It is worth noting
that at large distance $R\gg R_{eq}$, when the system separates to a
proton and a hydrogen atom, the potential surface exhibits a relative
maximum at $\sim 45^{\circ}$ inclination which is due to interaction
of the proton charge with the quadrupole moment of the
atom~\cite{Potekhin:01065402}.  As the sign of the interaction term is
angular dependent, a barrier is built up as the molecule is oriented
from parallel towards perpendicular
configuration~\cite{Turbiner:06309}.  With increasing field strength
and inclination, the internuclear distance at equilibrium becomes
smaller while the rotational barrier is increased. Data are presented
in Table~\ref{TableBTRE}. In Figure~\ref{h2pRT} the equilibrium
distance $R_{eq}(B,\theta)$ is plotted as a function of $\theta$ for
the field strengths $B = 0.1 \, B_0$ and $0.2\, B_0$.
\begin{table}[htb!]
\caption{Total energy at equilibrium distance as function of the field strength and inclination.}
\begin{tabular}{cccccccc}
\br
$B/B_0$ & $\theta$ & $R_{eq}/a_0$ & Energy/$E_h$ & 
$B/B_0$ & $\theta$ & $R_{eq}/a_0$ & Energy/$E_h$ \\
\mr
                $0$    &   $-$         & 1.9971  & -0.602625 &                        &               &         &          \\
                       &               &         &           &                        &               &         &          \\
\multirow{3}{*}{$0.1$} & $0^{\circ}$   & 1.9920  & -0.601029 & \multirow{3}{*}{$0.6$} & $0^{\circ}$   & 1.8705  & -0.550864\\
                       & $45^{\circ}$  & 1.9897  & -0.600785 &                        & $45^{\circ}$  & 1.8201  & -0.543923 \\
                       & $90^{\circ}$  & 1.9882  & -0.600613 &                        & $90^{\circ}$  & 1.7968  & -0.539131\\
                       &               &         &           &                        &               &         &          \\
\multirow{3}{*}{$0.2$} & $0^{\circ}$   & 1.9786  & -0.596311 & \multirow{3}{*}{$0.7$} & $0^{\circ}$   & 1.8399  & -0.534186\\
                       & $45^{\circ}$  & 1.9687  & -0.595361 &                        & $45^{\circ}$  & 1.7800  & -0.525296 \\
                       & $90^{\circ}$  & 1.9637  & -0.594678 &                        & $90^{\circ}$  & 1.7535  & -0.519216\\
                       &               &         &           &                        &               &         &          \\
\multirow{3}{*}{$0.3$} & $0^{\circ}$   & 1.9566  & -0.588667 & \multirow{3}{*}{$0.8$} & $0^{\circ}$   & 1.8096  & -0.515853\\
                       & $45^{\circ}$  & 1.9379  & -0.586615 &                        & $45^{\circ}$  & 1.7411  & -0.504917 \\
                       & $90^{\circ}$  & 1.9283  & -0.585161 &                        & $90^{\circ}$  & 1.7112  & -0.497503\\
                       &               &         &           &                        &               &         &          \\
\multirow{3}{*}{$0.4$} & $0^{\circ}$   & 1.9301  & -0.578360 & \multirow{3}{*}{$0.9$} & $0^{\circ}$   & 1.7799  & -0.496041\\ 
                       & $45^{\circ}$  & 1.9013  & -0.574889 &                        & $45^{\circ}$  & 1.7027  & -0.482994 \\
                       & $90^{\circ}$  & 1.8862  & -0.572447 &                        & $90^{\circ}$  & 1.6721  & -0.474219\\
                       &               &         &           &                        &               &         &            \\ 
%
\multirow{3}{*}{$0.5$} & $0^{\circ}$   & 1.9019  & -0.565667 & \multirow{3}{*}{$1.0$} & $0^{\circ}$   & 1.7563  & -0.474937\\
                       & $45^{\circ}$  & 1.8610  & -0.560550 &                         & $45^{\circ}$  & 1.6687  & -0.459670\\
                       & $90^{\circ}$  & 1.8413  & -0.556976 &                         & $90^{\circ}$  & 1.6348  & -0.449532\\
\end{tabular}
\label{TableBTRE}
\end{table}

\subsection{Magnetic Susceptibility}
An important quantity that describes the response of the molecular
system with respect to the external field is the magnetic
susceptibility. It is defined via a Taylor expansion of the
electronic energy in powers of the magnetic field $B$
\begin{equation} \label{etaylor}
  E(\vec{B})=  E(0)-\sum_{\alpha}c_{\alpha}B_{\alpha}-\frac{1}{2}\sum_{\alpha\beta}\chi_{\alpha\beta}B_{\alpha}B_{\beta}+\ldots
\end{equation}
For the electronic ground state, when the spin contributions are
neglected, the first coefficient, $c_\alpha$, vanishes.  The
coefficient tensor $\chi_{\alpha\beta}$ is the magnetic
susceptibility. The response of a molecule to an external magnetic
field leads to a classification into two types (see for example
\cite{Landau3}): diamagnetic and paramagnetic.

In the electronic Hamiltonian \eqref{ElectHamil} there are two terms
containing the magnetic field, $B$, a linear and a quadratic
one. Correspondingly, there are two contributions to the
susceptibility: a paramagnetic contribution originating from the
linear term of the Hamiltonian when treated by second order
perturbation theory in $B$, and a diamagnetic contribution coming from
the quadratic term in the first order in perturbation theory in $B$.

At first, let us proceed to the diamagnetic susceptibility.  The
diamagnetic susceptibility term $\chi^d$ can be expressed as the
expectation value with respect to the field-free wavefunction at
equilibrium distance.  Thus, in the symmetric gauge,
$\vec{A}_e=\frac{1}{2}\vec{B}\times\vec{\rm r}$, the expression of the
diamagnetic susceptibility tensor is
\begin{equation}
\chi_{\alpha\beta}^d= -\frac{1}{4}\left[\langle\vec{r}^{2}\rangle\delta_{\alpha\beta}-\langle r_{\alpha}r_{\beta}\rangle\right],
\end{equation}
where $\vec{r}=(x,y,z)$ is the position vector of the electron and
$r_{\alpha}$, $\alpha=1,2,3$, its components. If the magnetic field
direction is chosen along the $z$-axis, $\vec{B}=B\hat{z}$, the tensor
$\chi^d$ contains a single non-zero component, $\chi_{zz}^d \equiv
\chi^d$,
\begin{equation}
\label{chi1}
\chi^d=-\frac{1}{4}\langle\rho^2\rangle=-\frac{1}{4}\left[\langle x^2\rangle+\langle y^2\rangle\right]\,.
\end{equation}
Let us now consider the molecule in the $x$-$z$ plane (it can be
regarded as the definition of the $x$-direction).  For different
orientations of the molecule with respect to the \hbox{$z$-axis}, the
expectation values change according to a rotation by the angle
$\theta$ around the $y$-axis ($x\to x\cos\theta+z\sin\theta$, $y\to
y$, $z\to z\cos\theta-x\sin\theta$)
\begin{equation}
\label{chi2}
\chi^d=-\frac{1}{4}\left[\langle x^2\rangle_0\left(1+\cos^2\theta\right)+\langle z^2\rangle_0\sin^2\theta\right],
\end{equation}
where $\langle x^2\rangle_0$ and $\langle z^2\rangle_0$ are the
expectation values at zero inclination, and where we have used the
fact that $\langle x^2\rangle_0=\langle y^2\rangle_0$ and $\langle
xz\rangle_0=0$. In Table~\ref{Tablechi}, the numerical values of the
expectation values of the squares of the components of the position
vector of the electron, and the diamagnetic susceptibility, $\chi^d$,
are presented, at equilibrium distance, as function of $\theta$ and
compared with results obtained by Hegstrom~\cite{Hegstrom:7917} for
parallel and perpendicular orientations.
%
\begin{table}[htb]
\caption{\label{Tablechi}Expectation values of $x^2$, $y^2$ and $z^2$ for $B=0$, evaluated at the equilibrium distance
$R=R_{eq}=1.9971 \, a_0$, and diamagnetic, paramagnetic and total susceptibility ($\chi^d$, $\chi^p$, $\chi$) 
as function of $\theta$. $\chi^p$, obtained as $\chi^p=\chi-\chi^d$, is included for convenience.}
\begin{tabular}{cccclll}
\\
\br
$\theta$& $\langle x^2\rangle$&$\langle y^2\rangle$&$\langle z^2\rangle$&$\chi^d$&$\chi^p$&$\chi$\\
\mr
$0^{\circ}$                   &   0.64036  &  0.64036  &  1.11131  & -0.32018                   & 0.00000 & -0.32018\\
                              &            &           &           & -0.3209\cite{Hegstrom:7917}&  \phantom{0} ---       &   -0.3209\\
$15^{\circ}$                  &   0.67192  &  0.64037  &  1.07976  & -0.32807                   & 0.00022 & -0.32785\\
$30^{\circ}$                  &   0.75810  &  0.64035  &  0.99359  & -0.34961                   & 0.00216 & -0.34745\\
$45^{\circ}$                  &   0.87583  &  0.64040  &  0.87583  & -0.37906                   & 0.00992 & -0.36914\\
$60^{\circ}$                  &   0.99357  &  0.64038  &  0.75811  & -0.40849                   & 0.02062 & -0.38787\\
$75^{\circ}$                  &   1.07971  &  0.64042  &  0.67196  & -0.43003                   & 0.03090 & -0.39913\\
$90^{\circ}$                  &   1.11125  &  0.64041  &  0.64041  & -0.43792                   & 0.03447 & -0.40345\\
                              &            &           &           & -0.4382\cite{Hegstrom:7917}&   0.0378\cite{Hegstrom:7917}      &  -0.4004\cite{Hegstrom:7917} 
\end{tabular}
\end{table}

\begin{figure}[h!]
\centering
\includegraphics[angle=-90, width=0.85\textwidth]{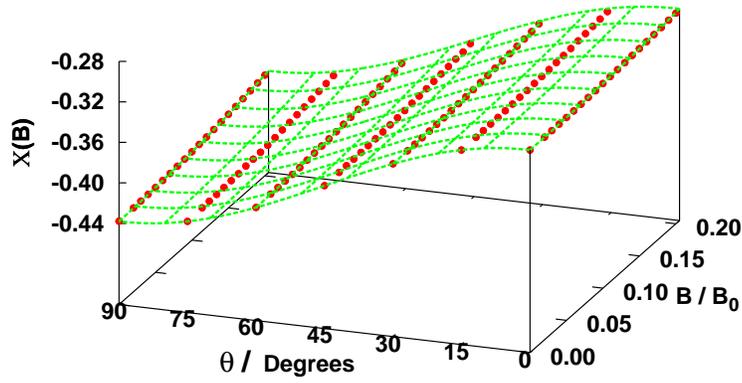}
\caption{Numerical values of the $X(B)=-\langle\rho^2\rangle_B/4$ (red
  dots) and the surface generated by \eqref{chia} as function of the
  magnetic field $\vec{B}$. The angle $\theta$ is the orientation of
  the molecule with respect to the field orientation.}
\label{Chisurface}
\end{figure}

\begin{figure}[h!]
\centering
\includegraphics[angle=-90, width=0.85\textwidth]{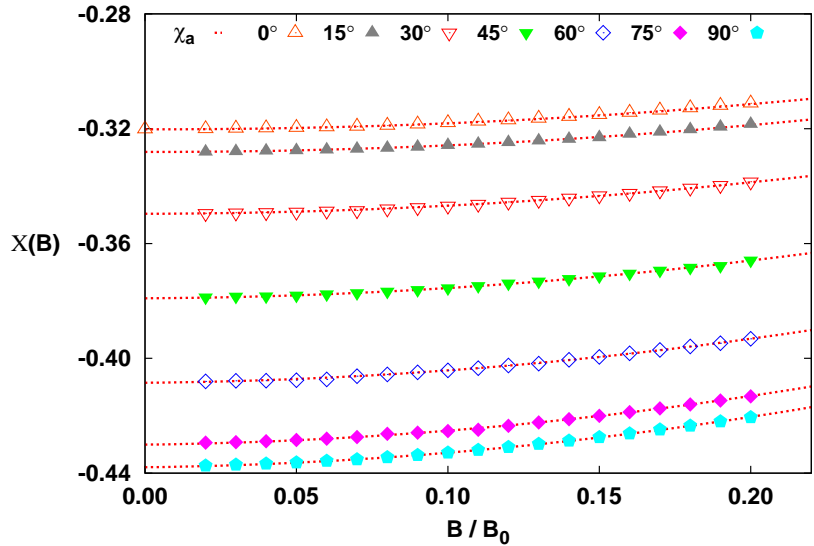}
\caption{Numerical values of the $X(B)=-\langle\rho^2\rangle_B/4$ and
  cuts of model \eqref{chia}, $X_a$, as function of the field strength
  $B$, for different given orientations.}
\label{Chicuts}
\end{figure}

For strong fields, higher powers of $B$ might need to be considered in
the expansion \eqref{etaylor}. With our variational method, evaluation
of higher order terms is straightforward as the trial functions depend
parametrically on the field strength. To this end we define the
function $X(\vec{B})=-\langle \rho^2\rangle_B/4$, where the
expectation value is taken with the optimized, $B$-dependent trial
function.  In the limiting case when $B\to0$, the diamagnetic
susceptibility is recovered, $X(\vec{B})\to\chi^d$.  Numerical results
of $X(\vec{B})$ were obtained at the equilibrium distances for $B=0.0,
0.01, \dots 0.2\, B_0$ and $\theta=0^{\circ}, 15^{\circ}, \dots
90^{\circ} $ and fitted ($rms = 1.6 \times 10^{-4}$) to a simple
functional form of the field strength $B$ and inclination $\theta$,
\begin{eqnarray}
\label{chia}
X_a(\vec{B})=&&-0.43795+0.013498B+0.37103B^2\\\nonumber
&&+\left(0.11774-0.01577B-0.141B^2\right)\cos^2\theta.
\end{eqnarray} 
This surface is plotted in Figure~\ref{Chisurface},
while cuts at constant orientation angles are presented in
Figure~\ref{Chicuts}.
It can be seen that the $X(\vec{B})$ is a smooth function of the
orientation angle, $\theta$, and the field strength, $B$, and tends to
the magnetic susceptibility $X(\vec{B})\to \chi^d$ as the field tends
to zero.  The points on the ordinate represent the value of the
magnetic susceptibility $\chi^d$ for various orientations and agree
with the corresponding data obtained from the model~\eqref{chia},
within the accuracy.  At weak fields $\lesssim $ $0.04\,B_0$, $X$ is
close to $\chi^d$ as given by \eqref{chi2}.  It indicates that
perturbation theory in $B$ can be applied and can provide sufficiently
accurate results. Eventually the diamagnetic susceptibility can be
given ($rms = 1.6 \times 10^{-4}$) as
\begin{equation}
\label{chia2}
\chi^d=-0.43795 + 0.11774 \cos^2\theta\,,
\end{equation} 
(c.f. (\ref{chia})).
Turning now to the total susceptibility $\chi$. In principle, it can
be obtained directly using the Taylor expansion~\eqref{etaylor} of the
energy potential curve $E=E(B,R,\theta)$ for fixed $\theta$ in powers
of $B$, but taken into account that the equilibrium distance
$R_{eq}(B,\theta)$ evolves in $B$. It is a quite complicated
procedure.  It is much easier to calculate numerically the energy
evolution with $B$ at minimum of the energy potential curve at fixed
inclination. Then interpolate this curve $E(B)$ near the origin, $B=0$
using a polynomial of finite degree in $B$.  The total susceptibility
$\chi$ will be related to the coefficient $E^{(2)}$ in front of the
$B^2$ term.  Numerical values of the total susceptibility, defined as
$\chi=-2E^{(2)}$ for different inclination $\theta$, are presented in
Table~\ref{Tablechi}.  They can be fitted accurately ($rms=3.45\times
10^{-4}$) to the following expression,
\begin{equation}
\label{chi3}
\chi=-0.41067+0.08260\cos^2{\theta}+0.007620\cos^2{2\theta}\,.
\end{equation}
Hence, for arbitrary inclination, the diamagnetic and total
susceptibility can be obtained using the expressions
(\ref{chia2}) and \eqref{chi3}.  Finally, the paramagnetic contribution
to the susceptibility can be evaluated as the difference $\chi^p=\chi
- \chi^d$, see data in Table~\ref{Tablechi}. In general, 
the paramagnetic susceptibility is much smaller 
that the diamagnetic part. It grows with inclination.

Concerning our statement that standard first order perturbation theory
based on the field-free $H^{(0)}$ problem should be applicable up to
$B \approx 0.04 \, B_0$, we may now add that at least 92\% of the
total susceptibility is recovered in this way. The energy correction
quadratic in $B$ is accurate to $\sim 6\times 10^{-5} \, E_h$.

\section{Solving the nuclear Schr\"odinger equation}

Substituting in the Hamiltonian~\eqref{AHamil} the electronic part
(last three terms) by the potential energy surface,
$\tilde{V}(R,\theta)$, we obtain the nuclear Hamiltonian. In the
symmetric gauge, it can be written as
\begin{equation}
\label{NucHamil}
\hat{\cal H}_{nuc}=\frac{2}{M_s}\hat{P}_R^2-\frac{1}{M_s}\vec{B}\cdot\hat{L}_R+\frac{1}{8M_s}\left[ B^2\vec{R}^2-(\vec{B}\cdot\vec{R})^2\right]\ +\tilde{V}(R,\theta) ,
\end{equation}
where $\hat{L}_R=\vec{R}\times\hat{P}_R$ is the angular momentum
operator of the molecular frame.  Transforming the Hamiltonian in
spherical coordinates yields
\begin{equation} \label{NucHamil2}
\hat{\cal H}_{nuc}=-\frac{2}{M_s}\frac{1}{R}\frac{\partial^2}{\partial R^2}R +\frac{2}{M_sR^2}\hat{L}_R^2-\frac{1}{M_s}B\hat{L}_z 
+\frac{1}{8M_s}B^2R^2\sin^2{\theta}+\tilde{V}(R,\theta),
\end{equation}
where $\hat{L}_z$ is the projection of angular momentum along $z$-axis
and $\theta$ the angle between the molecular and the $z$-axis.

We have solved the nuclear Schr\"odinger equation with
Hamiltonian~\eqref{NucHamil2} numerically. To this end the Hamiltonian
is divided in the two separate terms
\begin{equation}\label{H1}
\hat{\cal H}_{1}=-\frac{2}{M_s}\frac{1}{R}\frac{\partial^2}{\partial R^2}R 
+\tilde{V}(R,\theta^{\prime}) 
+\frac{1}{8M_s}B^2R^2\sin^2{\theta^{\prime}}
\end{equation}
and
\begin{equation}\label{H2}
\hat{\cal H}_{2}= \frac{2}{M_sR^2}\hat{L}_R^2-\frac{1}{M_s}B\hat{L}_z 
+\frac{1}{8M_s}B^2R^2\left(\sin^2\theta-\sin^2\theta^{\prime}\right)
+\tilde{V}(R,\theta)-\tilde{V}(R,\theta^{\prime})
\end{equation}
which roughly correspond to a vibrational part of the molecule with
reference orientation angle $\theta^{\prime}$, and a rotational part.
The rovibrational wave function is then expanded in terms of
vibrational and rotational basis functions as
\begin{equation}\label{rv_basis}
\Psi(R,\theta,\phi) = \sum_{v,L} c_{v,L} \frac{ \xi_{v}(R;\underline{\theta}^{\prime})}{R} 
Y_L^M(\theta,\phi)
\end{equation}
where $\xi_{v}(R;\underline{\theta}^{\prime})$ are solutions of
\eqref{H1}, obtained by numerical integration using the
renormalized Numerov algorithm, and $Y_L^M(\theta,\phi)$ are spherical
harmonics.  The matrix elements of Hamiltonian~\eqref{NucHamil2} in
this basis are
\begin{eqnarray}\label{Hnuc_mat1}
\left\langle v^{\prime} L^{\prime} M | \hat{\cal H}_{nuc} | v L M \right\rangle & = &
E_v(\theta^{\prime}) \delta_{L^{\prime} L}\delta_{v^{\prime} v}
+ \frac{2}{M_s} \left\langle v^{\prime} | \frac{1}{R^2} | v \right\rangle L(L+1) \delta_{L^{\prime}L} \nonumber \\
& - & \frac{B M }{M_s} \delta_{L^{\prime} L}\delta_{v^{\prime} v} \\ \nonumber
& + & \frac{B^2}{8M_s} \left\langle L^{\prime} M | \sin^2\theta-\sin^2\theta^{\prime} | LM \right\rangle 
\left\langle v^{\prime} | R^2 | v \right\rangle  \nonumber \\ 
& + & \left\langle v^{\prime} L^{\prime} M | \tilde{V}(R,\theta)-\tilde{V}(R,\theta^{\prime})
| v L M \right\rangle  \nonumber
\end{eqnarray}
with $E_v(\theta^{\prime})$ the eigenvalue of the vibrational
operator~\eqref{H1}.  To evaluate these matrix elements, the potential
$\tilde{V}(R,\theta)$ is presented as that of a hindered rotator,
\begin{eqnarray}\label{pot}
\tilde{V}(R,\theta) & = & \tilde{V}(R,0)
+ \sum_n \frac{V_{90,n}(R)}{2} \left[ 1 - \cos (2 n \theta) \right] \nonumber \\
& \approx & \tilde{V}(R,0) + V_{90}(R) \sin^2 \theta
\end{eqnarray}
%
$V_{90}(R)=\tilde{V}(R,90)-\tilde{V}(R,0)$ is the barrier height for
a given value of~$R$. 

\begin{figure}[h!]
\centering
\includegraphics[angle=-90, width=0.85\textwidth]{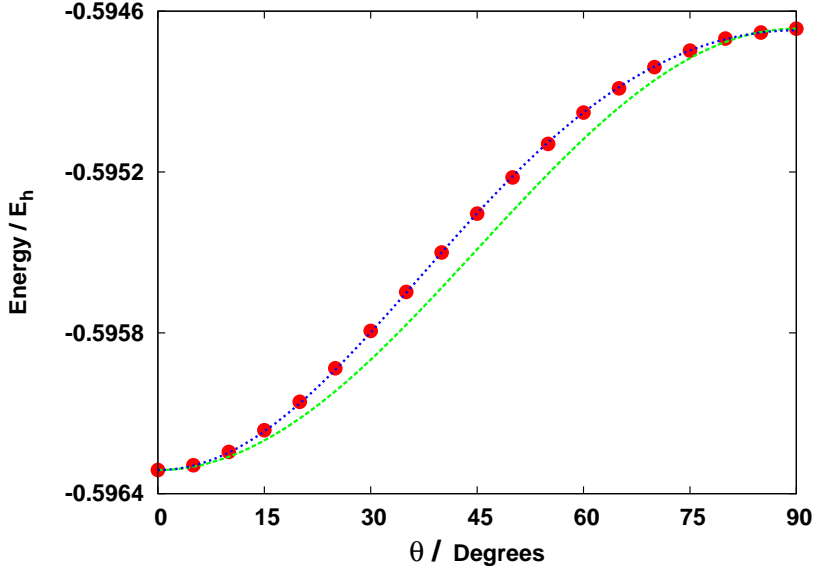}
\caption{Comparison between the hindered rotor approximation at first
  order (green line) and second order (blue) with a cut of the potential energy surface at
  $R_{eq}=1.9788 \, a_0$, the equilibrium distance when molecule 
  is oriented parallel to the field, for $B = 0.2 B_0$.}
\label{rotap}
\end{figure}

Limitation of the above expansion to just one term is a good
approximation of the potential at the field strengths considered in
the present work, as we have verified numerically. The $rms$ values
for fits of $V(R_{eq},\theta)$, using increments $\Delta \theta =
5^{\circ}$, are $rms=7.98\times 10^{-5}$ and $rms=3.66\times 10^{-6}$
for the one and two-term approximations and $B=0.2 B_0$. For $B=0.1
B_0$ the fitting error is reduced by a factor of four,
approximately. Figure~\ref{rotap} shows the performance of the two
approximations.
The one-term approximation thus represents the potential energy
surface to within the accuracy of the raw data of the electronic
energy.  An appealing feature is that just two slices, at
$\theta=0^{\circ}$ and $\theta=90^{\circ}$ of the surface are needed
explicitly.  Within the one-term approximation and choosing the
reference orientation $\theta^{\prime}=0$, the matrix elements can be
evaluated readily as~\footnote{ We use
  $\sin^2{\theta}=\frac{2}{3}-\frac{2}{3}\sqrt{\frac{4\pi}{5}}Y_2^0(\theta,\phi)$
  and the expression for the scalar product of three spherical
  harmonics, the Gaunt coefficients, $\int d\Omega
  Y_L^{M*}(\Omega)Y_{l_1}^{m_1}(\Omega)Y_{l_2}^{m_2}(\Omega)=(-1)^M\sqrt{\frac{(2l_1+1)(2l_2+1)(2L+1)}{4\pi}}
  \left(
\begin{array}{ccc}
l_1&l_2&L\\
0&0&0
\end{array}
\right)
\left(
\begin{array}{ccc}
l_1&l_2&L\\
m_1&m_2&-M
\end{array}
\right)$. The Condon-Shortly phase \cite{MagnascoBook} convention has been adopted.}
\begin{eqnarray}\label{Hnuc_mat2}
\left\langle v^{\prime} L^{\prime} M | \hat{\cal H}_{nuc} | v L M \right\rangle  & = &
E_v \delta_{L^{\prime} L}\delta_{v^{\prime} v} 
+ \frac{2}{M_s} \left\langle v^{\prime} | \frac{1}{R^2} | v \right\rangle L(L+1) \delta_{L^{\prime}L} \nonumber \\
& - & \frac{B M }{M_s} \delta_{L^{\prime} L}\delta_{v^{\prime} v}  \nonumber \\
& + & \left[\frac{B^2}{12M_s} \langle v^{\prime} | R^2 | v \rangle 
+\frac{2}{3}\langle v^{\prime} | V_{90}(R) | v \rangle \right] \delta_{L^{\prime}L} \nonumber \\
& - &\left[\frac{B^2}{12M_s} \langle v^{\prime} | R^2 | v \rangle 
+\frac{2}{3}\langle v^{\prime} | V_{90}(R) | v \rangle \right]  \nonumber \\ 
& \times & (-1)^{M}\sqrt{(2L'+1)(2L+1)}   \nonumber \\ 
& \times & \left(
\begin{array}{ccc}
L&2&L'\\
0&0&0
\end{array}
\right)
\left(
\begin{array}{ccc}
L&2&L'\\
M&0&-M
\end{array}
\right)
\end{eqnarray}
The terms in parentheses are Wigner $3j$-symbols. The
matrix~\eqref{Hnuc_mat2} is diagonal in $M$ as expected, since $M$ is
an exact quantum number. $L$-functions are coupled in steps of 2,
conserving parity.

\subsection{Results}

For the isotopologues \Htp and \Dtp we have computed the rovibrational
eigenvalues of the nuclear Hamiltonian for the four lowest vibrational
states and rotational excitation up to $L=5$ with respect to the
field-free case. Two levels of approximation are considered: a
simplified model in which only the diagonal terms with respect to the
vibrational basis are retained, and a second model which consists in
diagonalizing the Hamiltonian~\eqref{Hnuc_mat2} in the full basis.
These data are presented in Tables \ref{Table:rotv0} to
\ref{Table:Drotv3}.  The results obtained at the two levels of
approximation agree to within $10^{-5} \, E_h$.
%
%

\begin{table}
\caption{Rotational energy levels of \Htp in presence of a uniform magnetic field $B$ for the vibrational state $v=0$. The pure 
vibrational state ($L=0$ in the field-free case) is forbidden but shown here nevertheless as it corresponds to the origin of 
the rotational band. In the simple model 1, terms off-diagonal in $v$ are neglected. In model 2, the full
matrix~\eqref{Hnuc_mat2} is diagonalized.}
\label{Table:rotv0}
\centering
\begin{tabular}{ccrrcccc}
\\
$L$  & Energy/$E_h$        &$M$ &$\pi$&  \multicolumn{4}{c}{Energy/$E_h$} \\ \hline \hline
     & $B=0.0$ &    &     &  \multicolumn{2}{c}{$B=0.1$}&\multicolumn{2}{c}{$B=0.2$}\\
     &         &    &     & model 1 & model 2 & model 1 & model 2 \\\hline \hline
           &            &   -5 &    1 &  -0.591213 &  -0.591278 &  -0.584991 &  -0.585024 \\
           &            &    5 &    1 &  -0.591485 &  -0.591550 &  -0.585535 &  -0.585568 \\
           &            &   -4 &   -1 &  -0.591308 &  -0.591375 &  -0.585330 &  -0.585371 \\
           &            &    4 &   -1 &  -0.591525 &  -0.591593 &  -0.585765 &  -0.585807 \\
           &            &   -3 &    1 &  -0.591386 &  -0.591455 &  -0.585581 &  -0.585628 \\
$L=5$      & -0.593475  &    3 &    1 &  -0.591550 &  -0.591618 &  -0.585907 &  -0.585955 \\
           &            &   -2 &   -1 &  -0.591449 &  -0.591520 &  -0.585769 &  -0.585822 \\
           &            &    2 &   -1 &  -0.591558 &  -0.591629 &  -0.585987 &  -0.586040 \\
           &            &   -1 &    1 &  -0.591498 &  -0.591569 &  -0.585902 &  -0.585957 \\
           &            &    1 &    1 &  -0.591553 &  -0.591623 &  -0.586011 &  -0.586066 \\
           &            &    0 &   -1 &  -0.591532 &  -0.591604 &  -0.585982 &  -0.586038 \\
\hline
           &            &   -3 &    1 &  -0.593689 &  -0.593697 &  -0.587619 &  -0.587621 \\
           &            &    3 &    1 &  -0.593852 &  -0.593860 &  -0.587946 &  -0.587948 \\
           &            &   -2 &   -1 &  -0.593814 &  -0.593823 &  -0.588093 &  -0.588098 \\
$L=3$      & -0.595803  &    2 &   -1 &  -0.593923 &  -0.593932 &  -0.588311 &  -0.588315 \\
           &            &   -1 &    1 &  -0.593894 &  -0.593904 &  -0.588257 &  -0.588265 \\
           &            &    1 &    1 &  -0.593948 &  -0.593958 &  -0.588366 &  -0.588374 \\
           &            &    0 &   -1 &  -0.593938 &  -0.593948 &  -0.588408 &  -0.588417 \\
\hline
           &            &   -1 &    1 &  -0.595124 &  -0.595124 &  -0.589358 &  -0.589364 \\
$L=1$      & -0.597120  &    1 &    1 &  -0.595178 &  -0.595178 &  -0.589467 &  -0.589473 \\
           &            &    0 &   -1 &  -0.595327 &  -0.595327 &  -0.590107 &  -0.590109 \\
\hline
$L=0$      & -0.597386  &    0 &   -1 &  -0.595492 &  -0.595492 &  -0.590136 &  -0.590139 \\

\end{tabular}
\end{table}

\begin{table}
\caption{Rotational energy levels of \Htp in presence of a uniform magnetic field $B$ for the vibrational state $v=1$.
See Caption of Table~\ref{Table:rotv0} for explications.}
\label{Table:rotv1}
\centering
\begin{tabular}{ccrrcccc}
\\
$L$  & Energy/$E_h$        &$M$ &$\pi$&  \multicolumn{4}{c}{Energy/$E_h$} \\ \hline \hline
     & $B=0.0$ &    &     &  \multicolumn{2}{c}{$B=0.1$}&\multicolumn{2}{c}{$B=0.2$}\\
     &         &    &     & model 1 & model 2 & model 1 & model 2 \\\hline \hline
           &            &   -4 &    1 &  -0.582599 &  -0.582624 &  -0.576132 &  -0.576140 \\
           &            &    4 &    1 &  -0.582817 &  -0.582842 &  -0.576567 &  -0.576575 \\
           &            &   -3 &   -1 &  -0.582722 &  -0.582749 &  -0.576576 &  -0.576588 \\
           &            &    3 &   -1 &  -0.582885 &  -0.582912 &  -0.576903 &  -0.576915 \\
$L=4$      & 0.584909   &   -2 &    1 &  -0.582813 &  -0.582841 &  -0.576830 &  -0.576848 \\
           &            &    2 &    1 &  -0.582922 &  -0.582950 &  -0.577048 &  -0.577066 \\
           &            &   -1 &   -1 &  -0.582878 &  -0.582907 &  -0.577015 &  -0.577036 \\
           &            &    1 &   -1 &  -0.582932 &  -0.582961 &  -0.577124 &  -0.577145 \\
           &            &    0 &    1 &  -0.582917 &  -0.582946 &  -0.577103 &  -0.577124 \\
\hline
           &            &   -2 &    1 &  -0.584457 &  -0.584459 &  -0.578207 &  -0.578211 \\
           &            &    2 &    1 &  -0.584566 &  -0.584567 &  -0.578424 &  -0.578429 \\
$L=2$      & -0.586642  &   -1 &   -1 &  -0.584636 &  -0.584638 &  -0.578874 &  -0.578876 \\
           &            &    1 &   -1 &  -0.584690 &  -0.584692 &  -0.578982 &  -0.578985 \\
           &            &    0 &    1 &  -0.584679 &  -0.584681 &  -0.578674 &  -0.578684 \\
\hline
$L=0$      & -0.587395  &    0 &    1 &  -0.585400 &  -0.585401 &  -0.579854 &  -0.579857 \\

\end{tabular}
\end{table}

\begin{table}
\caption{Rotational energy levels of \Htp in presence of a uniform magnetic field $B$ for the vibrational state $v=2$. The pure vibrational state
  ($L=0$ in the field-free case) is forbidden but shown here nevertheless as it corresponds to the origin of the rotational band
See Caption of Table~\ref{Table:rotv0} for explications.}
\label{Table:rotv2}
\centering
\begin{tabular}{ccrrcccc}
\\
$L$  & Energy/$E_h$        &$M$ &$\pi$&  \multicolumn{4}{c}{Energy/$E_h$} \\ \hline \hline
     & $B=0.0$ &    &     &  \multicolumn{2}{c}{$B=0.1$}&\multicolumn{2}{c}{$B=0.2$}\\
     &         &    &     & model 1 & model 2 & model 1 & model 2 \\\hline \hline
           &            &   -5 &    1 &  -0.571973 &  -0.572030 &  -0.565079 &  -0.565103 \\
           &            &    5 &    1 &  -0.572245 &  -0.572302 &  -0.565623 &  -0.565648 \\
           &            &   -4 &   -1 &  -0.572091 &  -0.572151 &  -0.565505 &  -0.565536 \\
           &            &    4 &   -1 &  -0.572309 &  -0.572368 &  -0.565940 &  -0.565972 \\
           &            &   -3 &    1 &  -0.572186 &  -0.572248 &  -0.565798 &  -0.565836 \\
$L=5$      & -0.574480  &    3 &    1 &  -0.572349 &  -0.572411 &  -0.566125 &  -0.566163 \\
           &            &   -2 &   -1 &  -0.572261 &  -0.572324 &  -0.566017 &  -0.566061 \\
           &            &    2 &   -1 &  -0.572370 &  -0.572433 &  -0.566235 &  -0.566279 \\
           &            &   -1 &    1 &  -0.572316 &  -0.572380 &  -0.566165 &  -0.566212 \\
           &            &    1 &    1 &  -0.572371 &  -0.572435 &  -0.566274 &  -0.566321 \\
           &            &    0 &   -1 &  -0.572353 &  -0.572417 &  -0.566251 &  -0.566300 \\
\hline
           &            &   -3 &    1 &  -0.574209 &  -0.574215 &  -0.567495 &  -0.567498 \\
           &            &    3 &    1 &  -0.574372 &  -0.574379 &  -0.567822 &  -0.567825 \\
           &            &   -2 &   -1 &  -0.574368 &  -0.574376 &  -0.568099 &  -0.568102 \\
$L=3$      & -0.576565  &    2 &   -1 &  -0.574477 &  -0.574485 &  -0.568316 &  -0.568319 \\
           &            &   -1 &    1 &  -0.574461 &  -0.574470 &  -0.568200 &  -0.568212 \\
           &            &    1 &    1 &  -0.574515 &  -0.574524 &  -0.568309 &  -0.568321 \\
           &            &    0 &   -1 &  -0.574511 &  -0.574520 &  -0.568431 &  -0.568439 \\
\hline
           &            &   -1 &    1 &  -0.575526 &  -0.575526 &  -0.569257 &  -0.569261 \\
$L=1$      & -0.577746  &    1 &    1 &  -0.575581 &  -0.575581 &  -0.569366 &  -0.569370 \\
           &            &    0 &   -1 &  -0.575789 &  -0.575789 &  -0.570150 &  -0.570151 \\
\hline
$L=0$      & -0.577984  &    0 &   -1 &  -0.575902 &  -0.575903 &  -0.570160 &  -0.570161 \\

\end{tabular}
\end{table}

\begin{table}
\caption{Rotational energy levels of \Htp in presence of a uniform magnetic field $B$ for the vibrational state $v=3$.
See Caption of Table~\ref{Table:rotv0} for explications.} 
\label{Table:rotv3}
\centering
\begin{tabular}{ccrrcccc}
\\
$L$  & Energy/$E_h$        &$M$ &$\pi$&  \multicolumn{4}{c}{Energy/$E_h$} \\ \hline \hline
     & $B=0.0$ &    &     &  \multicolumn{2}{c}{$B=0.1$}&\multicolumn{2}{c}{$B=0.2$}\\
     &         &    &     & model 1 & model 2 & model 1 & model 2 \\\hline \hline
           &            &   -4 &    1 &  -0.564365 &  -0.564387 &  -0.557218 &  -0.557224 \\
           &            &    4 &    1 &  -0.564583 &  -0.564605 &  -0.557654 &  -0.557660 \\
           &            &   -3 &   -1 &  -0.564514 &  -0.564538 &  -0.557775 &  -0.557783 \\
           &            &    3 &   -1 &  -0.564678 &  -0.564702 &  -0.558102 &  -0.558110 \\
$L=4$      & -0.566912  &   -2 &    1 &  -0.564621 &  -0.564647 &  -0.558029 &  -0.558045 \\
           &            &    2 &    1 &  -0.564730 &  -0.564756 &  -0.558247 &  -0.558263 \\
           &            &   -1 &   -1 &  -0.564695 &  -0.564721 &  -0.558257 &  -0.558275 \\
           &            &    1 &   -1 &  -0.564749 &  -0.564775 &  -0.558366 &  -0.558384 \\
           &            &    0 &    1 &  -0.564737 &  -0.564764 &  -0.558332 &  -0.558353 \\
\hline
           &            &   -2 &    1 &  -0.566047 &  -0.566048 &  -0.559197 &  -0.559201 \\
           &            &    2 &    1 &  -0.566156 &  -0.566157 &  -0.559415 &  -0.559419 \\
$L=2$      & -0.568461  &   -1 &   -1 &  -0.566268 &  -0.566270 &  -0.560019 &  -0.560019 \\
           &            &    1 &   -1 &  -0.566323 &  -0.566324 &  -0.560128 &  -0.560128 \\
           &            &    0 &    1 &  -0.566297 &  -0.566300 &  -0.559640 &  -0.559653 \\
\hline
$L=0$      & -0.569134  &    0 &    1 &  -0.566962 &  -0.566962 &  -0.561037 &  -0.561037 \\

\end{tabular}
\end{table}


\begin{table}
\caption{Rotational energy levels of \Dtp in presence of a uniform magnetic field $B$ for the vibrational state $v=0$.
See Caption of Table~\ref{Table:rotv0} for explications.} 
\label{Table:Drotv0}
\centering
\begin{tabular}{ccrrcccc}
\\
$L$  & Energy/$E_h$        &$M$ &$\pi$&  \multicolumn{4}{c}{Energy/$E_h$} \\ \hline \hline
     & $B=0.0$ &    &     &  \multicolumn{2}{c}{$B=0.1$}&\multicolumn{2}{c}{$B=0.2$}\\
     &         &    &     & model 1 & model 2 & model 1 & model 2 \\\hline \hline
           &            &   -4 &    1 &  -0.595506 &  -0.595511 &  -0.589494 &  -0.589496 \\
           &            &    4 &    1 &  -0.595615 &  -0.595620 &  -0.589712 &  -0.589714 \\
           &            &   -3 &   -1 &  -0.595598 &  -0.595604 &  -0.589871 &  -0.589874 \\
           &            &    3 &   -1 &  -0.595680 &  -0.595686 &  -0.590035 &  -0.590037 \\
$L=4$      & -0.597576  &   -2 &    1 &  -0.595663 &  -0.595669 &  -0.590013 &  -0.590019 \\
           &            &    2 &    1 &  -0.595717 &  -0.595724 &  -0.590122 &  -0.590128 \\
           &            &   -1 &   -1 &  -0.595706 &  -0.595713 &  -0.590166 &  -0.590173 \\
           &            &    1 &   -1 &  -0.595733 &  -0.595740 &  -0.590221 &  -0.590227 \\
           &            &    0 &    1 &  -0.595729 &  -0.595736 &  -0.590197 &  -0.590205 \\
\hline
           &            &   -2 &    1 &  -0.596502 &  -0.596502 &  -0.590702 &  -0.590706 \\
           &            &    2 &    1 &  -0.596557 &  -0.596557 &  -0.590811 &  -0.590815 \\
$L=2$      & -0.598506  &   -1 &   -1 &  -0.596641 &  -0.596641 &  -0.591252 &  -0.591254 \\
           &            &    1 &   -1 &  -0.596668 &  -0.596669 &  -0.591306 &  -0.591308 \\
           &            &    0 &    1 &  -0.596652 &  -0.596653 &  -0.591894 &  -0.591896 \\
\hline
$L=0$      & -0.598908  &    0 &    1 &  -0.597052 &  -0.597052 &  -0.590953 &  -0.590961 \\

\end{tabular}
\end{table}

\begin{table}
\caption{Rotational energy levels of \Dtp in presence of a uniform magnetic field $B$ for the vibrational state $v=1$.
See Caption of Table~\ref{Table:rotv0} for explications.} 
\label{Table:Drotv1}
\centering
\begin{tabular}{ccrrcccc}
\\
$L$  & Energy/$E_h$        &$M$ &$\pi$&  \multicolumn{4}{c}{Energy/$E_h$} \\ \hline \hline
     & $B=0.0$ &    &     &  \multicolumn{2}{c}{$B=0.1$}&\multicolumn{2}{c}{$B=0.2$}\\
     &         &    &     & model 1 & model 2 & model 1 & model 2 \\\hline \hline
           &            &   -5 &    1 &  -0.587598 &  -0.587610 &  -0.581316 &  -0.581319 \\
           &            &    5 &    1 &  -0.587734 &  -0.587746 &  -0.581589 &  -0.581592 \\
           &            &   -4 &   -1 &  -0.587688 &  -0.587701 &  -0.581666 &  -0.581671 \\
           &            &    4 &   -1 &  -0.587797 &  -0.587810 &  -0.581884 &  -0.581889 \\
           &            &   -3 &    1 &  -0.587757 &  -0.587772 &  -0.581865 &  -0.581874 \\
$L=5$      & -0.589802  &    3 &    1 &  -0.587839 &  -0.587854 &  -0.582029 &  -0.582037 \\
           &            &   -2 &   -1 &  -0.587809 &  -0.587825 &  -0.582027 &  -0.582037 \\
           &            &    2 &   -1 &  -0.587864 &  -0.587879 &  -0.582136 &  -0.582146 \\
           &            &   -1 &    1 &  -0.587846 &  -0.587862 &  -0.582121 &  -0.582133 \\
           &            &    1 &    1 &  -0.587873 &  -0.587889 &  -0.582175 &  -0.582187 \\
           &            &    0 &   -1 &  -0.587867 &  -0.587883 &  -0.582176 &  -0.582189 \\
\hline
           &            &   -3 &    1 &  -0.588812 &  -0.588813 &  -0.582680 &  -0.582683 \\
           &            &    3 &    1 &  -0.588894 &  -0.588895 &  -0.582844 &  -0.582847 \\
           &            &   -2 &   -1 &  -0.588938 &  -0.588939 &  -0.583179 &  -0.583181 \\
$L=3$      & -0.590947  &    2 &   -1 &  -0.588992 &  -0.588994 &  -0.583288 &  -0.583290 \\
           &            &   -1 &    1 &  -0.588999 &  -0.589001 &  -0.583114 &  -0.583122 \\
           &            &    1 &    1 &  -0.589026 &  -0.589028 &  -0.583168 &  -0.583177 \\
           &            &    0 &   -1 &  -0.589033 &  -0.589035 &  -0.583374 &  -0.583379 \\
\hline
           &            &   -1 &    1 &  -0.589548 &  -0.589548 &  -0.583868 &  -0.583870 \\
$L=1$      & -0.591589  &    1 &    1 &  -0.589575 &  -0.589575 &  -0.583922 &  -0.583925 \\
           &            &    0 &   -1 &  -0.589757 &  -0.589757 &  -0.584512 &  -0.584513 \\
\hline
$L=0$      & -0.591718  &    0 &   -1 &  -0.589795 &  -0.589796 &  -0.583503 &  -0.583511 \\

\end{tabular}
\end{table}

\begin{table}
\caption{Rotational energy levels of \Dtp in presence of a uniform magnetic field $B$ for the vibrational state $v=2$.
See Caption of Table~\ref{Table:rotv0} for explications.} 
\label{Table:Drotv2}
\centering
\begin{tabular}{ccrrcccc}
\\
$L$  & Energy/$E_h$        &$M$ &$\pi$&  \multicolumn{4}{c}{Energy/$E_h$} \\ \hline \hline
     & $B=0.0$ &    &     &  \multicolumn{2}{c}{$B=0.1$}&\multicolumn{2}{c}{$B=0.2$}\\
     &         &    &     & model 1 & model 2 & model 1 & model 2 \\\hline \hline
           &            &   -4 &    1 &  -0.581345 &  -0.581350 &  -0.574886 &  -0.574889 \\
           &            &    4 &    1 &  -0.581454 &  -0.581459 &  -0.575104 &  -0.575106 \\
           &            &   -3 &   -1 &  -0.581458 &  -0.581463 &  -0.575341 &  -0.575343 \\
           &            &    3 &   -1 &  -0.581539 &  -0.581545 &  -0.575505 &  -0.575507 \\
$L=4$      & -0.583590  &   -2 &    1 &  -0.581533 &  -0.581539 &  -0.575442 &  -0.575450 \\
           &            &    2 &    1 &  -0.581587 &  -0.581593 &  -0.575551 &  -0.575559 \\
           &            &   -1 &   -1 &  -0.581582 &  -0.581588 &  -0.575649 &  -0.575656 \\
           &            &    1 &   -1 &  -0.581609 &  -0.581616 &  -0.575703 &  -0.575710 \\
           &            &    0 &    1 &  -0.581607 &  -0.581614 &  -0.575643 &  -0.575654 \\
\hline
           &            &   -2 &    1 &  -0.582282 &  -0.582282 &  -0.576120 &  -0.576124 \\
           &            &    2 &    1 &  -0.582336 &  -0.582336 &  -0.576229 &  -0.576233 \\
$L=2$      & -0.584451  &   -1 &   -1 &  -0.582453 &  -0.582453 &  -0.576748 &  -0.576749 \\
           &            &    1 &   -1 &  -0.582480 &  -0.582480 &  -0.576803 &  -0.576804 \\
           &            &    0 &    1 &  -0.582446 &  -0.582447 &  -0.577428 &  -0.577429 \\
\hline
$L=0$      &  -0.584823 &    0 &    1 &  -0.582847 &  -0.582848 &  -0.576348 &  -0.576357 \\

\end{tabular}
\end{table}

\begin{table}
\caption{Rotational energy levels of \Dtp in presence of a uniform magnetic field $B$ for the vibrational state $v=3$.
See Caption of Table~\ref{Table:rotv0} for explications.} 
\label{Table:Drotv3}
\centering
\begin{tabular}{ccrrcccc}
\\
$L$  & Energy/$E_h$        &$M$ &$\pi$&  \multicolumn{4}{c}{Energy/$E_h$} \\ \hline \hline
     & $B=0.0$ &    &     &  \multicolumn{2}{c}{$B=0.1$}&\multicolumn{2}{c}{$B=0.2$}\\
     &         &    &     & model 1 & model 2 & model 1 & model 2 \\\hline \hline
           &            &   -5 &    1 &  -0.574069 &  -0.574083 &  -0.567312 &  -0.567315 \\
           &            &    5 &    1 &  -0.574205 &  -0.574219 &  -0.567585 &  -0.567588 \\
           &            &   -4 &   -1 &  -0.574176 &  -0.574190 &  -0.567733 &  -0.567736 \\
           &            &    4 &   -1 &  -0.574285 &  -0.574299 &  -0.567951 &  -0.567954 \\
           &            &   -3 &    1 &  -0.574257 &  -0.574271 &  -0.567927 &  -0.567935 \\
$L=5$      & -0.576443  &    3 &    1 &  -0.574338 &  -0.574353 &  -0.568090 &  -0.568099 \\
           &            &   -2 &   -1 &  -0.574316 &  -0.574331 &  -0.568120 &  -0.568130 \\
           &            &    2 &   -1 &  -0.574370 &  -0.574386 &  -0.568229 &  -0.568239 \\
           &            &   -1 &    1 &  -0.574356 &  -0.574372 &  -0.568210 &  -0.568223 \\
           &            &    1 &    1 &  -0.574384 &  -0.574399 &  -0.568265 &  -0.568277 \\
           &            &    0 &   -1 &  -0.574379 &  -0.574394 &  -0.568278 &  -0.568291 \\
\hline
           &            &   -3 &    1 &  -0.575203 &  -0.575205 &  -0.568648 &  -0.568651 \\
           &            &    3 &    1 &  -0.575285 &  -0.575286 &  -0.568811 &  -0.568814 \\
           &            &   -2 &   -1 &  -0.575354 &  -0.575355 &  -0.569236 &  -0.569236 \\
$L=3$      & -0.577502  &    2 &   -1 &  -0.575408 &  -0.575410 &  -0.569345 &  -0.569345 \\
           &            &   -1 &    1 &  -0.575417 &  -0.575419 &  -0.569941 &  -0.569942 \\
           &            &    1 &    1 &  -0.575444 &  -0.575446 &  -0.569996 &  -0.569997 \\
           &            &    0 &   -1 &  -0.575457 &  -0.575459 &  -0.569426 &  -0.569430 \\
\hline
           &            &   -1 &    1 &  -0.575911 &  -0.575911 &  -0.569061 &  -0.569071 \\
$L=1$      & -0.578097  &    1 &    1 &  -0.575938 &  -0.575938 &  -0.569115 &  -0.569126 \\
           &            &    0 &   -1 &  -0.576160 &  -0.576160 &  -0.570634 &  -0.570634 \\
\hline
$L=0$      & -0.578216  &    0 &   -1 &  -0.576183 &  -0.576183 &  -0.569491 &  -0.569498 \\

\end{tabular}
\end{table}

\begin{figure}[htp!]
\centering
\includegraphics*[angle=-90, width=0.85\textwidth]{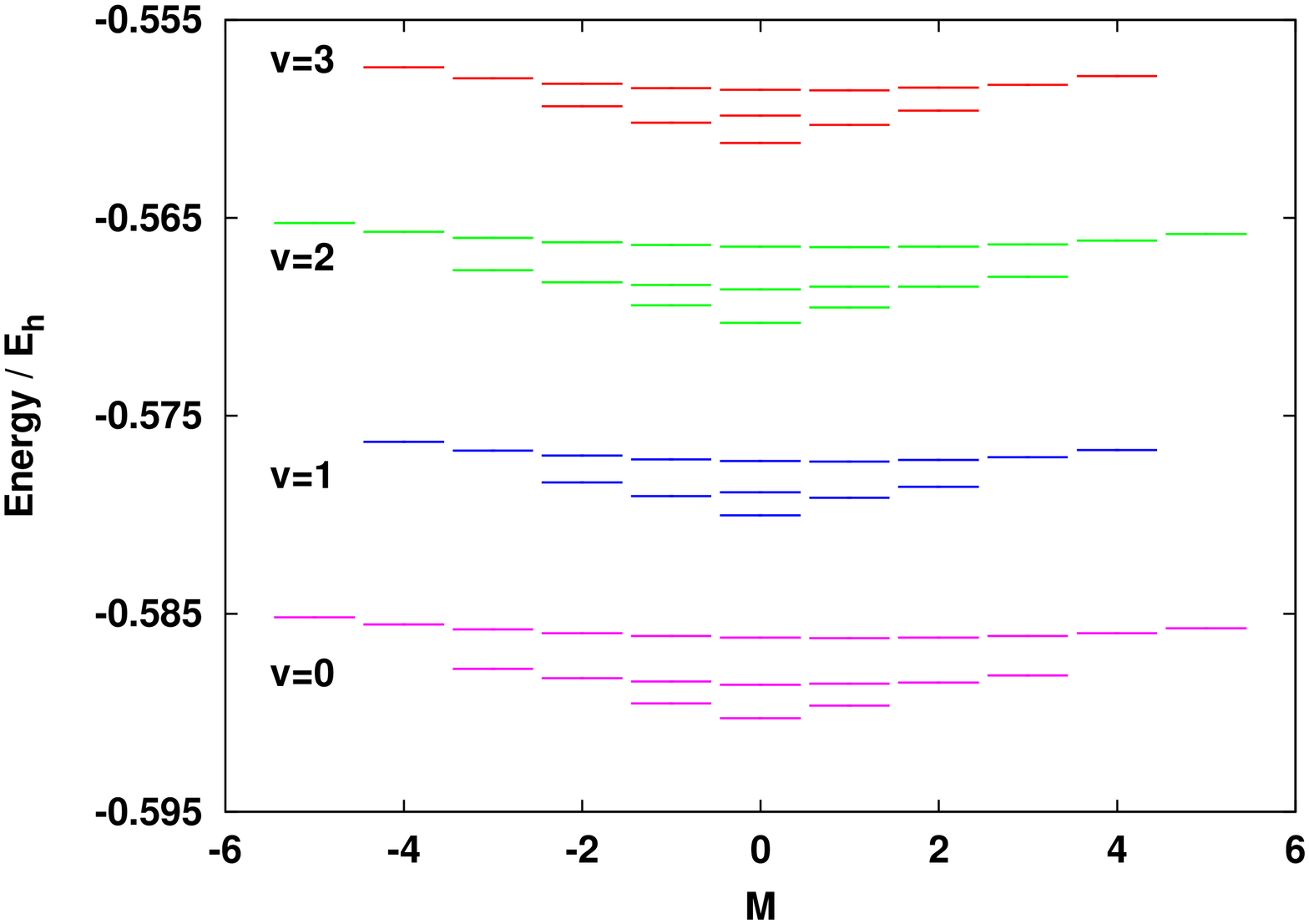}
\caption{Rotational structure, up to $L=5$, of the four lowest vibrational
states of \Htp in the presence of an external magnetic field of $B=0.2 B_0$.
All states are located above the rotational barrier, which is at $E= -0.595678 \; E_h$.
}
\label{rovibplot}
\end{figure}

If spin effects are neglected, rovibrational states of \Htp in a
magnetic field can be classified in terms of three quantum numbers:
the vibrational quantum number, $\nu$, the projection of the angular
momentum of the molecular frame on the field axis, $M$, and the
$z$-parity, $\pi$. The latter quantum number is due to the fact that
positive and negative $z$-directions of the field are equivalent. If
the wave function is reflected at the plane $z=0$, $\theta$ is mapped
to $\pi - \theta$ and $Y_L^M(\pi-\theta,\phi) = (-1)^{L+M}
Y_L^M(\theta,\phi)$.  The $z$-parity of the state is thus
$\pi=(-1)^{L+M}$. The nuclear wave function of a system of two
fermions must be antisymmetric with respect to an exchange of the
nuclei. The vibrational part of the wave function is symmetric for
even vibrational quanta, $v=0,2, \dots$ and antisymmetric for odd,
$v=1,3,\dots$.  The symmetry of the rotational part can be derived
from the properties of the spherical harmonics with respect to
inversion, ($\theta, \phi \rightarrow \theta+\pi, \phi+\pi$),
$Y_L^M(\theta+\pi,\phi+\pi) = (-1)^{L} Y_L^M(\theta,\phi)$.  Hence for
even $v$ the rotational functions must have odd parity, while for odd
$v$ they must have even parity, just as in the field-free case. The
expression for the $z$-parity is
\begin{equation} \label{zpar}
\pi = (-1)^{M+v+1} = \left\{ 
\begin{array}{rl}
-(-1)^M  & {\rm for} ~v~ {\rm even} \\
(-1)^M  & {\rm for} ~v~ {\rm odd} 
\end{array}
\right.
\end{equation}
For \Dtp, a system with two bosonic nuclei, vibrational and rotational
parts of the wavefunction must have the same parity. In this case, the
$z$-parity is given by
\begin{equation} \label{Dzpar}
\pi = (-1)^{M+v} = \left\{ 
\begin{array}{rl}
(-1)^M  & {\rm for} ~v~ {\rm even} \\
-(-1)^M  & {\rm for} ~v~ {\rm odd} 
\end{array}
\right.
\end{equation}
The calculated rovibrational states in Tables \ref{Table:rotv0} to
\ref{Table:Drotv3} are labeled with the exact quantum
numbers. Graphical analysis of the \Htp rovibrational states,
Figure~\ref{rovibplot}, shows that they remain grouped according to
the field-free quantum number $L$ which is explained by the fact that
all rovibrational states are located above the rotational barrier, for
the two isotopologues. $B=0.2 B_0$, or 47000 Tesla, is a strong field
but of modest size in atomic units, hence $L$ may be considered a good
quantum number. The main effect of the magnetic field in this region
of field strengths is on the electronic energy.

\clearpage

\section{Conclusions}

We have investigated the problem of \Htp and \Dtp in an external
magnetic field of up to $B = 0.2 B_0$ or 4.7$\times 10^4$ T by exact
and approximate methods.
This includes a thorough analysis of the electronic energy as function
of field strength and orientation of the molecule with respect to the
external field as well as of the rovibrational structure of \Htp and
\Dtp. The electronic problem has been solved by the variational method
with physically adequate trial functions.  It is shown that both
diamagnetic and paramagnetic susceptibilities grow with inclination,
while paramagnetic susceptibility is systematically much smaller than
the diamagnetic one.  Evaluation of the magnetic susceptibility shows
that first-order perturbation theory based on zero-field trial
functions may no longer be accurate at a field strength of above $B
\approx 0.04 B_0$. To solve the ro-vibrational problem, the hindered
rotor approximation, in which the potential energy surface is
approximated by a zero-inclination potential curve as a function of
the internuclear distance $R$ and a simple parametrization of the
rotational barrier in the angular coordinate $\theta$ gives results
accurate to about $10^{-5} \; E_h$, which is comparable in accuracy
with the rigid-rotor approximation to separate vibrational and
rotational motion. Some of the approximations have been used before,
by other authors, at much higher field strengths, were they are less
accurate.  The findings of the present paper provide a basis for
future investigations dealing with higher field strengths and
different molecules such as $\rm H_2$.

\section*{Acknowledgements}
H.M.C. is grateful to Consejo Nacional de Ciencia y Tecnolog\'{\i}a,
Mexico, for a postdoctoral grant (CONACyT grant no 202139). This work
was also supported by a Mexican-French binational research grant
(CONACyT-CNRS grant no 26218) and by the Computer Center ROMEO of the
University of Reims Champagne-Ardenne.  Research by J.C.L.V and
A.V. T. was supported in part by CONACyT grant 116189 and DGAPA
IN109512.
\newline

\end{document}